\documentclass[english,aps,prl,amsmath,twocolumn,preprintnumber,superscriptaddress,notitlepage]{revtex4-2}
\usepackage[T1]{fontenc}
\usepackage[latin9]{inputenc}
\makeatletter

\usepackage{mathptmx,newtxtext,newtxmath,xspace}
\usepackage{amsbsy,bm,bbold}
\usepackage{graphicx,color,xcolor,epsfig,rotate}
\usepackage{fancyhdr}
\usepackage[colorlinks=true, 
            linkcolor=blue, 
            urlcolor=blue,
           citecolor=blue]{hyperref}
\usepackage{soul}

\makeatother

\usepackage{babel}

\begin{document}
\title{Multipolar Skyrmion Crystals in Non-Kramers Doublet Systems}
\author{Hao~Zhang}
\affiliation{Theoretical Division and CNLS, Los Alamos National Laboratory, Los Alamos, New Mexico 87545, USA}
\author{Shi-Zeng~Lin}
\affiliation{Theoretical Division and CNLS, Los Alamos National Laboratory, Los Alamos, New Mexico 87545, USA}
\affiliation{Center for Integrated Nanotechnologies (CINT), Los Alamos National Laboratory, Los Alamos, New Mexico 87545, USA
}
\date{\today}
\begin{abstract}
We study the Kondo lattice model of multipolar magnetic moments interacting with conduction electrons on a triangular lattice. Bond-dependent electron hoppings induce a compass-like anisotropy in the effective Ruderman-Kittel-Kasuya-Yosida interaction between multipolar moments. This unique anisotropy stabilizes multipolar skyrmion crystals at zero magnetic field. In a unit cell, the skyrmion fractionalizes into meron composites subject to the conservation of total topological charge. Diverse multipolar phases in the phase diagram give rise to novel spontaneous Hall response of conduction electrons.
\end{abstract}
\maketitle

Magnetic skyrmion crystal (SkX) configurations were first theoretically proposed~\cite{bogolubskaya1989stationary,bogolubskaya1990stationary,polyakov1975metastable} and then experimentally observed~\cite{muhlbauer2009skyrmion, yu2010realspace,yu2011near,seki2012observation,adams2012longwavelength} in chiral magnets. SkX formations have also been found in centrosymmetric magnets stabilized by competing exchange or dipolar interactions~\cite{okubo2012multiple,leonov2015multiply,lin2016ginzburg,hayami2016bubble,batista2016frustration,wang_skyrmions_2020,hayami2021review,yu2012bfsmo,yu2014biskyrmion,mallik1998paramana,saha1999magnetic,kurumaji2019skyrmion,chandragiri2016magnetic,hirschberger2019}. Due to their topological nature and potential applications in spintronics~\cite{tokura2021magnetic}, magnetic skyrmions have been a central focus in condensed matter physics over the past decade. The skyrmions are typically based on a Kramers doublet and are commonly referred to as dipolar skyrmions. Here, we explore the emergence of SkXs within non-Kramers doublet systems~\cite{mueller1968effective}. In these systems, all constituents of the skyrmions are multipolar moments devoid of any dipolar character. The distinct symmetry properties of these underlying multipolar moments suggest that non-Kramers skyrmions may harbor novel physics that is not shared by their dipolar counterparts.

The non-Kramers doublet as ground state is common in both partially filled $d$ and $f$ shells with an even number of electrons, driven by the interplay between the crystal electric field (CEF) and spin-orbit coupling (SOC)~\cite{santini_multipolar_2009}. In this study, we focus on the latter scenario, with the aim of establishing connections with the family of intermetallic compounds $\mathrm{Pr(Ti,V,Tr)_2(Al,Zn)_{20}}$~\cite{sato2012ferroquad,onimaru2009supers,onimaru2011antiferroquad,onimaru2012simul}. Here, the interactions among multipolar moments that reside in the non-Kramers doublet subspace are mediated by conduction electrons, which is similar to the Ruderman-Kittel-Kasuya-Yosida (RKKY) interaction for the dipolar moments~\cite{ruderman1954,kasuya1956theory,yoshida1957}.

\begin{figure}
    \centering
    \includegraphics[width=1.0\columnwidth]{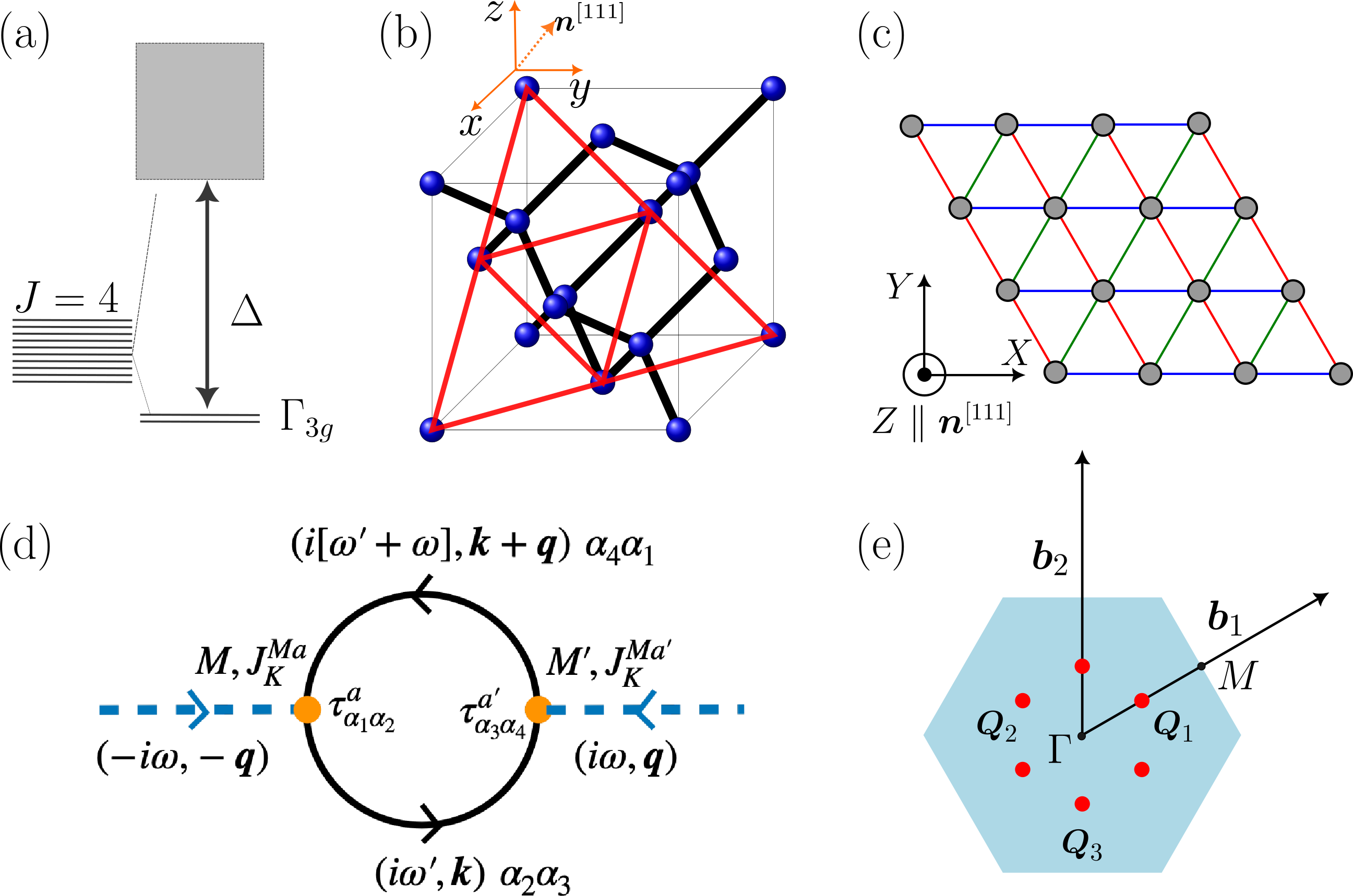}
    \caption{(a) Under $T_d$ CEF, degenerate $J=4$ levels of $5f^2$ electrons split, leading to the low-lying non-Kramers doublet $\Gamma_{3g}$ and a CEF gap $\Delta$. (b) Diamond lattice formed by $\mathrm{Pr}^{3+}$ ions (black circles) that host the $5f^2$ electrons. The [111] plane is a 2D triangular lattice (red lines). The orange dot line indicates the 3-fold axis $\bm{n}^{[111]}$. (c) Bond-dependent hoppings and pseudo-spin interactions on the triangular lattice. For the blue, red, and green bond, $\phi_{\gamma}=0$, $2\pi/3$, and $4\pi/3$, respectively. (d) The bubble diagram used to compute the magnetic susceptibility $\chi^{MM'}(\bm{q})$. (e) The red dots indicate the peak positions of $\chi^{MM'}(\bm{q})$ in the first BZ (shaded area) of the TL.} 
    \label{fig:1}
\end{figure}

The RKKY interaction is known to stabilize chiral dipolar orders such as skyrmion crystals~\cite{batista2008itinerant,motome2012hidden,wang_skyrmions_2020}. Like in typical skyrmion hosting systems~\cite{yu2012magnetic,okubo2012multiple,leonov2015multiply,hayami2016bubble,lin2016ginzburg}, achieving the SkX phase generally requires the presence of an external magnetic field.~\footnote{Zero-field SkXs have been reported to exist in a Kondo lattice model on the triangular lattice~\cite{ozawa2017zerofield} but in the strong coupling regime. Here we focus on the weak coupling regime, i.e., the RKKY limit.} However, the coupling between the magnetic field and the multipolar moments are of higher order~\cite{patri2019unveiling}. This implies that obtaining a multipolar SkX may necessitate the application of an unrealistically strong magnetic field. We show that our system, featuring a compass-like anisotropy originating from the non-Kramers nature of the constituent doublet and bond-dependent hoppings, stabilizes multipolar SkXs even at \emph{zero magnetic field}. Moreover, our model exhibits various multipolar phases. 
The absence of dipolar characteristics also poses a challenge for detecting multipolar orders, as they evade conventional magnetic probes such as neutron scattering and nuclear magnetic resonance~\cite{patri2019unveiling,lee2018landau}. Consequently, these orders are often referred to as ``hidden orders.''  We propose that the spontaneous Hall response of conduction electrons in the presence of these multipolar phases can serve as signature for detecting these orders.

We start by considering the single-ion physics for the $5f^2$ configurations. Under the local $T_d$ CEF, the spin-orbit coupled $J=4$ multiplet splits and gives rise to a low-lying non-Kramers doublet $\Gamma_{3g}$~\cite{patri2020critical}, separated from the high-energy levels by a CEF gap $\Delta$ [see Fig.~\ref{fig:1}~(a)]. Within the $\Gamma_{3g}$, three pseudo-spin-1/2 operators, defined as
\begin{equation}
    S^{X} \equiv -\frac{1}{16} (3 J_z^2 - \bm{J}^2),    \
    S^{Y} \equiv -\frac{\sqrt{3}}{16} (J_x^2 - J_y^2), \
    S^{Z} \equiv  \frac{\sqrt{3}}{36} \overline{J_x J_y J_z}
    \label{eq:ops}
\end{equation}
form a closed $\mathfrak{su}(2)$ algebra. Here $S^{X,Y}$ are time-reversal even quadrupolar operators, while $S^{Z}$~\footnote{The overline indicates the fully symmetric product} is a time-reversal odd octupolar operator.

\begin{figure*}[ht!]
    \centering
    \includegraphics[width=2.0\columnwidth]{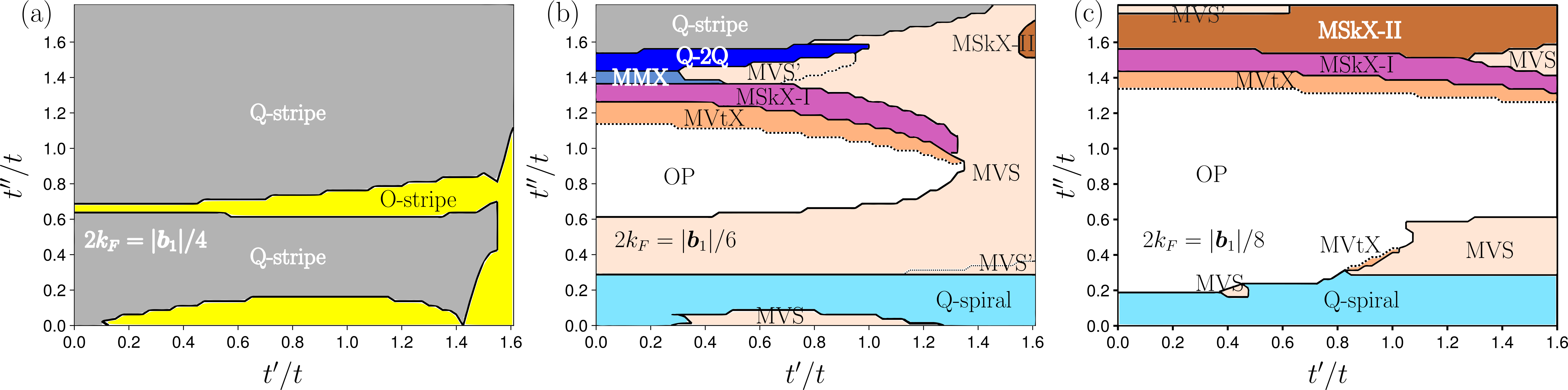}
    \caption{Phase diagrams for the multipolar RKKY model as functions of the anisotropic hopping amplitudes $t'/t$ and $t''/t$. Solid lines represent first-order phase boundaries, whereas dotted lines denote second-order phase boundaries. (a) $2k_F=\lvert \bm{b}_1 \rvert/4$ with filling $n_c=0.03-0.13$; (b) $2k_F=\lvert \bm{b}_1 \rvert/6$ with filling $n_c=0.016-0.035$; (c) $2k_F=\lvert \bm{b}_1 \rvert/8$ with filling $n_c=0.009-0.019$.} 
    \label{fig:2}
\end{figure*}

In general, the pseudo-spin operators $S^{X,Y,Z}$ interact with both the spin and orbital degrees of freedom of conduction electrons. In this work we focus on the simplest $e_g$ conduction electrons to illustrate our ideas. One important observation is that the symmetry-allowed Kondo interactions are always singlet for \emph{physical spins} of conduction electrons~\cite{patri_emergent_2020,patri2020critical,supp}. As a result, we suppress the physical spin indices, and the Kondo Hamiltonian is written as
\begin{equation}
    \mathcal{H}_{K} = J_{K_1} \sum_{j} \left[ \bm{c}_j^{\dagger} \left(S_j^X \tau^x - S_j^Y \tau^z \right) \bm{c}_j\right]
                    + J_{K_2} \sum_{j} S_j^Z \bm{c}_j^{\dagger} \tau^y \bm{c}_j,
    \label{eq:kondo}
\end{equation}
where $\bm{c}_j^{\dagger} = (c_{j1}^{\dagger}, c_{j2}^{\dagger})$ denotes the two-component spinor that carries the $e_g$ orbital indices, and $\tau$s are the Pauli matrices acting in this space.

Note that the two $e_g$ orbitals also form a non-Kramers doublet. We can also perform symmetry analysis to constrain the symmetry-allowed hoppings. We consider the nearest-neighbor (NN) tight-binding model for $e_g$ electrons on a 2D triangular lattice (TL), that can be viewed as the [111] plane of on the diamond lattice [see Fig.~\ref{fig:1}~(b)]. In the presence of the inversion, time-reversal and the three-fold rotation symmetry along the [111] direction, the Hamiltonian can be written as [see Fig.~\ref{fig:1}~(c)]
\begin{align}
    \mathcal{H}_{ij}^{\gamma; t} & = -t\bm{c}_{i}^{\dagger}\tau^{0}\bm{c}_{j} -t'\bm{c}_{i}^{\dagger}(i\tau^{y})\bm{c}_{j} \nonumber \\
                              & - t''\bm{c}_{i}^{\dagger}(\cos\phi_{\gamma}\tau^{z}-\sin\phi_{\gamma}\tau^{x})\bm{c}_{j}+h.c.,
    \label{eq:hopping}
\end{align}
where all hopping amplitudes are real. Here $\phi_\gamma =  0, 2\pi/3, 4\pi/4$ describes the bond-dependent hopping, see Fig.~\ref{fig:1} (c).

The full Kondo lattice problem is then described by the Kondo coupling Eq.~\eqref{eq:kondo} and the bond-dependent hopping Hamiltonian Eq.~\eqref{eq:hopping}. We focus on the weak-coupling limit $\lvert J_{K_{1,2}}/t \rvert \ll 1$, and take the classical limit for the multipolar moments, i.e., replacing the multipolar operators with their expectation values in the SU(2) coherent states formed by the $\Gamma_{3g}$ doublet~\cite{hao2021classical}. In this limit, the Kondo coupling is a perturbative effect relative to the Fermi energy of conduction electron,
we may integrate out conduction electrons to obtain a ``pseudo-spin-only'' Hamiltonian via the second-order perturbation theory (also known as the RKKY Hamiltonian):
\begin{equation}
    \mathcal{H}^{\text{RKKY}} = \sum_{\bm{q}} \chi^{MM'}(\bm{q})S_{-\bm{q}}^{M}S_{\bm{q}}^{M'}.
    \label{eq:rkky}
\end{equation}
Here $\chi^{MM'}(\bm{q}) \ (M, M' = X, Y, Z)$ is the magnetic susceptibility tensor, which can be obtained by evaluating the diagram depicted in Fig.~\ref{fig:1}(d) and subsequently taking the static limit:
\begin{align}
    \chi^{MM'}(\bm{q}) & = J_{K}^{Ma}J_{K}^{M'a'}\sum_{\bm{k}\in BZ}\sum_{n,n'=\pm} \frac{f(\varepsilon_{\bm{k}+\bm{q},n})-f(\varepsilon_{\bm{k},n'})}{\varepsilon_{\bm{k}+\bm{q},n}-\varepsilon_{\bm{k},n'}} \nonumber \\
                       &\times \mathrm{Tr}\big[\tau^{a}\hat{\mathcal{M}}_{\bm{k},n'}\tau^{a'}\hat{\mathcal{M}}_{\bm{k}+\bm{q},n}\big],
    \label{eq:susceptibility}
\end{align}
where $f(\varepsilon_{\bm{k},n})$ is the Fermi distribution function and the matrix element $[\mathcal{M}_{\bm{q}, n}]^{\alpha \alpha'} \equiv \mathcal{U}_{\alpha,n}(\bm{q})\mathcal{U}_{\alpha',n}^{*}(\bm{q})$ with $\mathcal{U}(\bm{q})$ the eigenvector matrix and of Eq.~\eqref{eq:hopping}~\cite{supp}. Here $J_K^{Ma}$ indicates the Kondo coupling between the multipolar component $M$ and the orbital Pauli matrix component $a$ [c.f. Eq.~\eqref{eq:kondo}] and the Greek letters $\alpha$s denote the $e_g$ orbital indices.

Now we discuss the dependence of $\chi^{MM'}(\bm{q})$ on the parameters $J_{K_{1,2}}$, $t'$, and $t''$ (we set $t\equiv1$ as the unit). First, the Kondo interactions $J_{K_1}$ and $J_{K_2}$ control relative strengths for the quadrupolar ($S^{X}, S^{Y}$) and the octupolar ($S^{Z}$) sectors, i.e., $J_{K_1} \neq J_{K_{2}}$ induces an ``XXZ-like'' RKKY interactions. As the Kondo interactions are just multiplicative factors for the RKKY Hamiltonian, and given that both $t'$ and $t''$ can also induce XXZ anisotropy, we set $J_{K_1}=J_{K_2}$ in the rest of this manuscript to simplify our discussions. If the hopping is isotropic and bond-independent ($t'=t''=0$), the tensor $\chi^{MM'}(\bm{q})$ becomes proportional to the identity matrix and the RKKY interaction is SU(2) invariant (within the pseudo-spin space), recovering the results given in Ref.~\cite{wang_skyrmions_2020} for dipolar spins. For finite $t'$ and $t''$, after Fourier transforming Eq.~\eqref{eq:rkky} to real space, the pseudo-spin interaction on the $\gamma$-bond [see Fig.~\ref{fig:1}~(c) for NN bonds as an example]~\footnote{Note that the RKKY interaction is long-ranged. Further neighbor interactions also take the same form but not shown in Fig.~\ref{fig:1}~(c)} takes the form
\begin{equation}
    \mathcal{H}_{ij}^{\gamma; \text{RKKY}} = J^{xy}_{ij}(S_{i}^{X}S_{j}^{X}+S_{i}^{Y}S_{j}^{Y})+J^{zz}_{ij}S_{i}^{Z}S_{j}^{Z}
    +J^{\gamma}_{ij} \tilde{S}_{i}^{\gamma} \tilde{S}_{j}^{\gamma},
    \label{eq:realspace}
\end{equation}
where $\tilde{S}_{i}^{\gamma} = \cos\phi_{\gamma} S_i^X + \sin\phi_{\gamma} S_i^Y$ is the in-plane pseudo-spin component along the $\gamma$-bond direction that manifests the three-fold rotation symmetry of the model. Couplings between the quadrupolar and octupolar moments like $S^{X(Y)} S^Z$ are absent because they violate the time-reversal symmetry.

We refer the last (bond-dependent) term of Eq.~\eqref{eq:realspace} as the ``compass-like'' term. This particular compass term has been demonstrated to favor the formation of six-sublattice ``vortex''-like multipolar textures in honeycomb and triangular lattice~\cite{chaloupka2015hidden,khaliullin2021exchange}. Additionally, on the square lattice, the associated compass-like anisotropy has been shown to stabilize skyrmion crystals at zero magnetic field~\cite{wang2021meron}. Here the strength $J^{\gamma}_{ij}$ is sorely determined by the ratio $t''/t$ [c.f. the bond-dependent hopping term in Eq.~\eqref{eq:hopping}] when fixing $J_{K_{1,2}}$. We notice that pseudo-spin Hamiltonians, presented in forms equivalent to Eq.~\eqref{eq:realspace}, have been explored in Refs.~\cite{liu2018selective,khaliullin2021exchange}. However, our derivation highlights the natural emergence of Eq.~\eqref{eq:realspace} as the low energy effective model from a microscopically more realistic high-energy model Eqs.~\eqref{eq:kondo}--\eqref{eq:hopping}. More importantly, the RKKY origin of Eq.~\eqref{eq:realspace}, which favors the formation of multipolar ordered states with ordering wave vector $\lvert \bm{Q}_{\nu} \rvert = 2k_F$~\cite{wang_skyrmions_2020}, inherently introduces a length scale bigger than the atomic lattice parameter. This length scale, in combination with the compass-like term, stabilizes the multipolar skyrmion crystals to be discussed below.

We compute the $T=0$ phase diagrams by numerically minimizing the \emph{classical} multipolar RKKY Hamiltonian Eq.~\eqref{eq:rkky} (see~\cite{supp} for details). The three phase diagrams in Fig.~\ref{fig:2} are obtained by fixing Fermi wave vectors at $2k_F=\lvert \bm{b}_1 \rvert/4, \lvert \bm{b}_1 \rvert/6, \lvert \bm{b}_1 \rvert/8$. 

\begin{figure}[htp!]
    \centering
    \includegraphics[width=1.0\columnwidth]{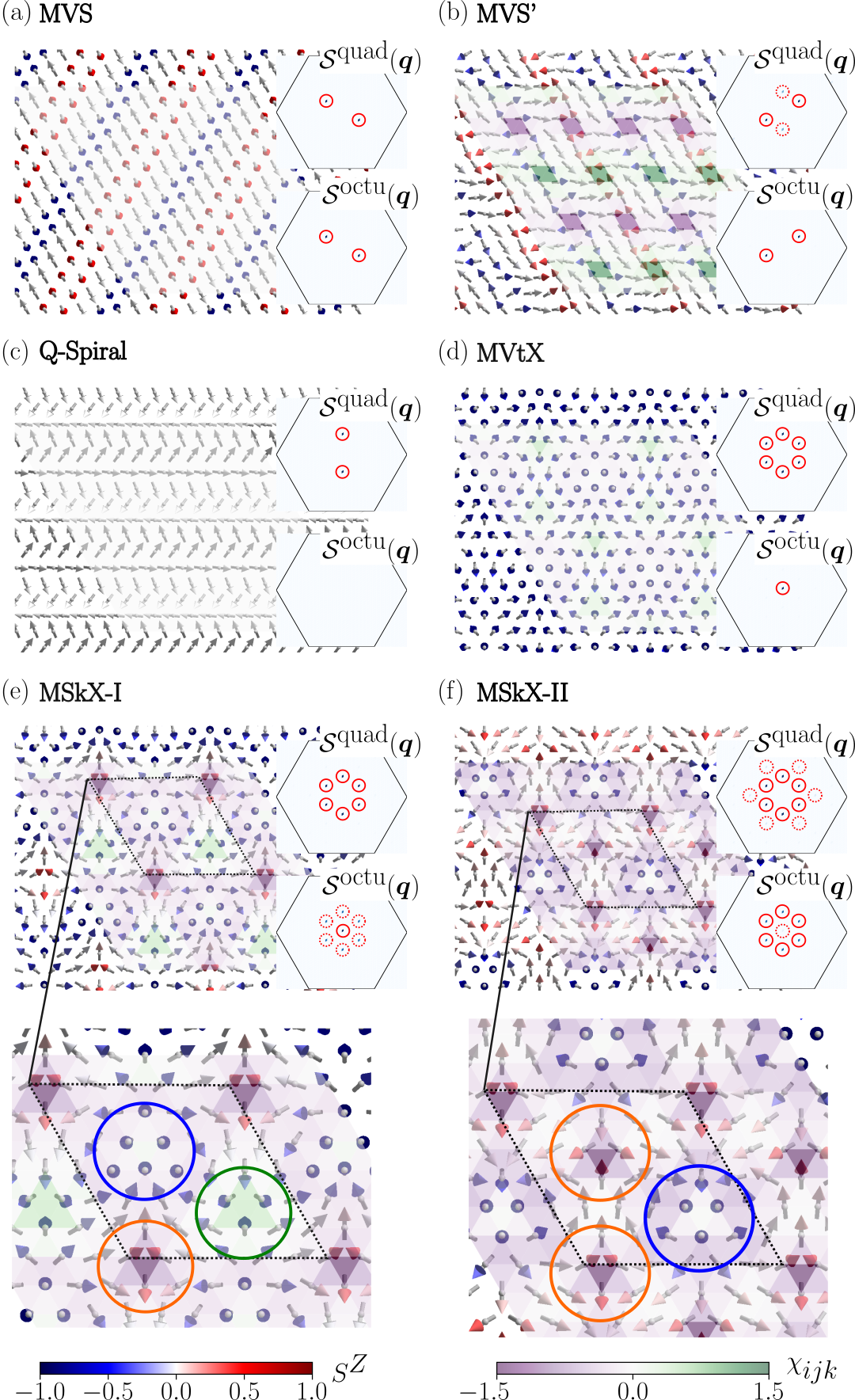}
    \caption{Multipolar (pseudo) spin configurations of some common phases of the three phase diagrams shown in Fig.~\ref{fig:2}. The in-plane components, $X$ and $Y$, represent the expectation values of two quadrupolar operators, while the out-of-plane component, $Z$, corresponds to the expectation of the octupolar component. Insets show the quadrupolar ($\mathcal{S}^{\text{quad}}$) and octupolar ($\mathcal{S}^{\text{octu}}$) structure factors in the first Brillouin Zone (BZ). Solid red circles emphasize dominant peaks, while dotted red circles indicate subdominant peaks. The panels below (e) and (f) present zoomed-in views of MSkX-I and MSkX-II within a single magnetic unit cell (indicated by the dotted parallelogram). The orange circle marks a meron  with a winding number $\mathcal{W}=-1$,  the blue circle denotes an anti-meron with $\mathcal{W}=2$, and the green circle indicates another anti-meron with $\mathcal{W}=-1$.}
    \label{fig:3}
\end{figure}

The phase diagrams shown in Fig.~\ref{fig:2} include a variety of multipolar phases. Within the multipolar vertical spiral (MVS) phase [Fig.~\ref{fig:3}(a)], the pseudo-spin expectation values trace a spiral pattern on a plane parallel the ``direction'' of the $Z$-component (octupolar) of the pseudo spin. The multipolar vertical spiral with the modulated quadrupolar components (MVS') phase depicted in Fig.~\ref{fig:3}(b) is related to the MVS phase through a second-order transition. Moving on to the Q-spiral phase presented in Fig.~\ref{fig:3}(c), it showcases an ``in-plane'' spiral exclusively formed by the quadrupolar components. The octupolar-polarized (OP) phase signifies the presence of ``ferromagnetic'' octupolar ordered states, establishing a connection to the multipolar vortex crystal (MVtX) phase through a second-order transition.  Following this OP to MVtX transition, the quadrupolar components organize into a 3-$\bm{Q}$ structure with vortex-like structure in a unit cell, while all octupolar moments maintain alignment in the same direction, leading to a zero skyrmion charge $n_{\text{sk}}\equiv \int dr^2 \mathbf{S}\cdot \partial_x\mathbf{S}\times\partial_y \mathbf{S}/4\pi=0$. Noteworthy are two emergent multipolar skyrmion crystals (MSkXs) for $2k_F=\lvert \bm{b}_1 \rvert/6, \lvert \bm{b}_1 \rvert/8$ [Fig.~\ref{fig:2}(b)(c)]. Both MSkXs phases feature a 3-$\bm{Q}$ structure in both quadrupolar and octupolar components. They differ from each other by the net skyrmion charge in a magnetic unit cell: MSkX-I has $n_{\text{sk}}=\pm 1$, whereas $n_{\text{sk}}=\pm 2$ for MSkX-II. 

The internal multipolar structure associated with skyrmion charges can be determined by examining the product of the vorticity (winding number of the quadrupolar components $\mathcal{W}\equiv \oint \nabla \phi\cdot d\mathbf{l}/2\pi$, where $\phi$ is the in-plane phase of quadrupolar components) and the polarization of the core moment (sign of the octupolar component) for each constituent~\cite{PhysRevB.91.224407,yu2018transformation,gao2019creation}. The two panels below Fig.~\ref{fig:3}(e) and (f) show zoomed-in views of MSkX-I and MSkX-II within a single magnetic unit cell.
MSkX-I comprises a meron with a winding number $\mathcal{W}=-1$, an anti-meron with $\mathcal{W}=2$, and another anti-meron with $\mathcal{W}=-1$. Consequently, the skyrmion charge for MSkX-I per unit cell is reproduced as $n_{\text{sk}}=\frac{1}{2}(-1) - \frac{1}{2}(2) - \frac{1}{2}(-1) = -1$, whereas each unit cell of MSkX-II consists of two merons with $\mathcal{W}=-1$ and one anti-meron with $\mathcal{W}=2$, resulting in  $n_{\text{sk}}=\frac{1}{2}(-1)\times 2 - \frac{1}{2}(2) = -2$. Therefore, MSkX-I and MSkX-II can also be viewed as a crystal of meron composites. The MSkXs with $\pm$ topological charge are degenerate due to the inversion or time reversal symmetry of Eq.~\eqref{eq:rkky}, but the system automatically chooses one flavor in the phase diagram as a result of the spontaneous symmetry breaking.  The remaining phases, including the quadrupolar stripe (Q-stripe), octupolar stripe (O-stripe), multipolar meron crystal (MMX), and quadrupolar double-$\bm{Q}$ (Q-2Q), are comprehensively described in the SM~\cite{supp}.

We proceed to provide understanding of the phase diagrams illustrated in Fig.~\ref{fig:2}. Each phase diagram corresponds to a specific range of conduction electron filling fractions, denoted as $n_c$. Maintaining a fixed $2k_F$ and adjusting $t'/t$ within the range of 0 to 1.6 and $t''/t$ within the range of 0 to 1.8, the variations in $n_c$ remain relatively modest. However, when $t'(t'')/t$ falls outside these specified ranges, a significant shift in $n_c$ occurs, and the peak locations of $\chi^{MM'}$ no longer necessarily reside along the $\Gamma \rightarrow M$ direction, as depicted in Fig.~\ref{fig:1}(e). Consequently, we constrain $t'/t$ and $t''/t$ to be within the mentioned ranges. A common feature across all three phase diagrams is that the phases exhibit more similarity with varying $t'/t$ but greater divergence with varying $t''/t$. This discrepancy arises because, in the small $n_c$ limit considered here, the \emph{bare} electron dispersion relation (see~\cite{supp}) can be expanded near the $\Gamma$-point as $\varepsilon_{\bm{q},\pm} = -6t + \left[ 3t/2 \pm 3t''/4\right]q^2 + \mathcal{O}(q^3)$. As a result, $t'/t$ must be significant to make substantial contributions.

When examining the two phase diagrams depicted in Fig.~\ref{fig:2}(b)(c), a similarity emerges upon rescaling both axes, implying a universal underlying physics that governs the stabilization of these phases. We point out that while the RKKY interaction introduces a length scale in the system, it is not the exclusive mechanism for stabilizing these textures. A short-ranged model with frustration can also establish a length scale, albeit with a less clear microscopic origin compared to the RKKY interaction. In SM~\cite{supp}, we begin with the frustrated $J_1$-$J_2$ version of the multipolar spin model (Eq.~\eqref{eq:rkky}), from which we derive the universal Ginzburg-Landau theory to elucidate this point. Here, we provide a brief summary of the results.
For small $J^{\gamma}$ [corresponding to small $t''/t$ in Fig.~\ref{fig:2}(b)(c)], the system evolves into an in-plane quadrupolar spiral (Q-spiral) or a multipolar vertex spiral (MVS), depending on whether the XXZ interaction is easy-plane or easy-axis.  If the easy-axis interaction is  dominant, the system polarizes fully into the OP phase. A significant $J^{\gamma}>0$ then promotes the vortex pattern, giving rise to the MVtX phase. As $J^{\gamma}$ increases further, a SkX-I emerges by leveraging  more in-plane quadrupolar components. The SkX-II always occurs at a larger $J^{\gamma}$ ($t''/t$) compared to SkX-I due to its additional in-plane components. Eventually, for a large enough $J^{\gamma}$, the system favors in-plane orders such as the Q-2Q and Q-stripe phases. The phase diagram shown in Fig.~\ref{fig:2}(a), corresponding to a larger electron filling $n_c$ and a shorter wavelength, exhibits greater dissimilarity because the aforementioned long-wavelength analysis no longer applies.

\begin{figure}
    \centering
    \includegraphics[width=1.0\columnwidth]{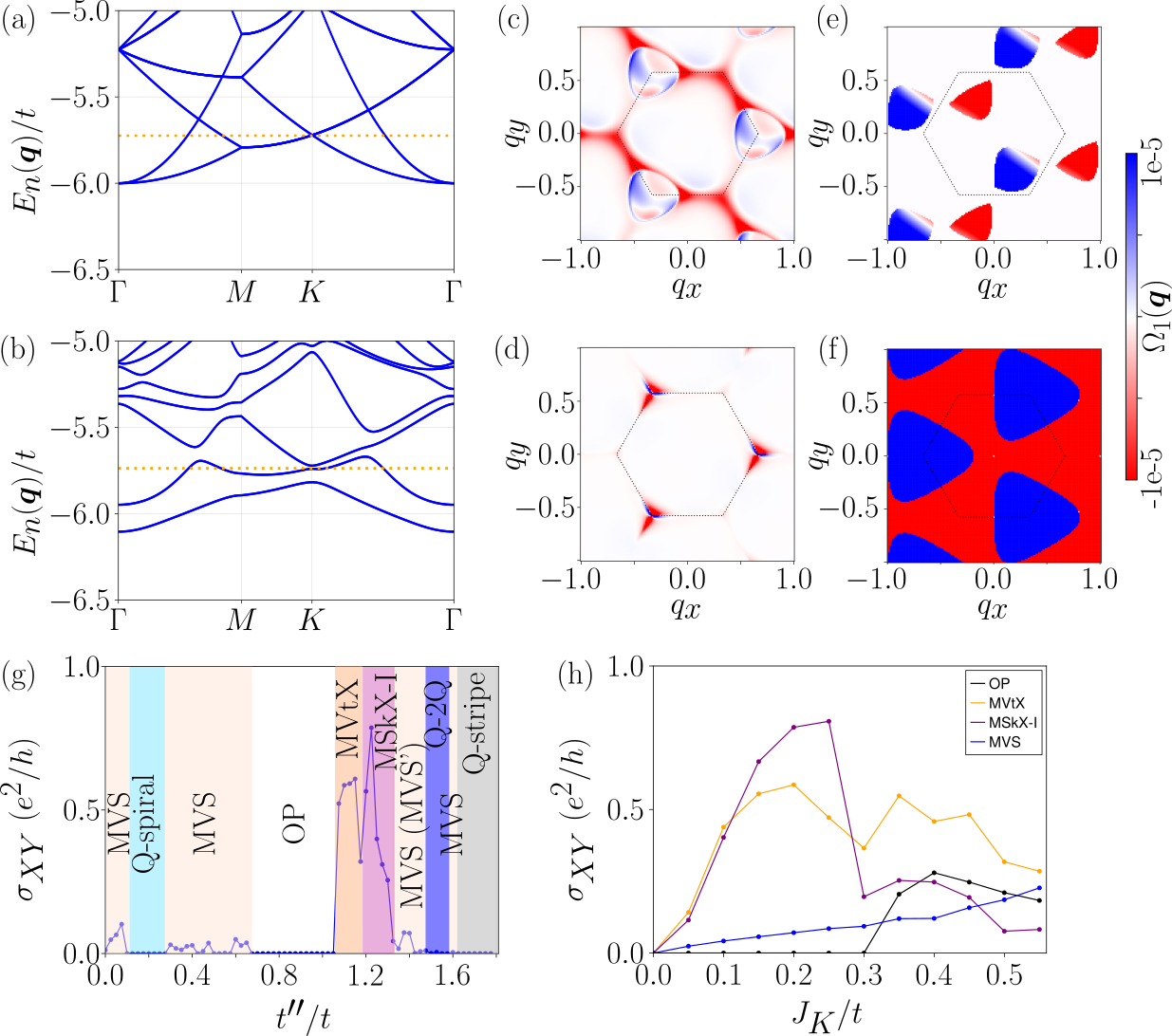}
    \caption{Electronic band structures for $t'/t=0.8$, $t''/t=1.225$: (a) $J_{K_{1,2}}=0$ and (b) $J_{K_{1}}=J_{K_2}=0.2t$. The orange dot lines in both (a) and (b) represent the Fermi levels, for a common filling fraction $n_c \approx 0.02055$. The \emph{bare} Fermi level in (a) corresponds to $2k_F=\lvert \bm{b}_1 \rvert/6$. (c)--(f) The Berry curvatures for the lowest band in momentum space for different phases under the conditions $t'/t=0.8$, $2k_F=\lvert \bm{b}_1 \rvert/6$, and $J_{K_{1}}=J_{K_2}=0.2t$: (c) MSkX-I ($t''/t=1.225$), (d) MVtX ($t''/t=1.15$), (e) MVS ($t''/t=1.4$), and (f) OP ($t''/t=0.8$). (g) Dependence of the Hall conductance $\sigma_{XY}$ on $t''/t$ while maintaining $t'/t=0.8$, $J_{K_{1}}=J_{K_2}=0.2t$, and $2k_F=\lvert \bm{b}_1 \rvert/6$. (h) The Hall conductance $\sigma_{XY}$ as a function of $J_{K_1}=J_{K_2}=J_K$ for the four phases represented in (c)--(f), with a constant $n_c$ for each phase: MSkX-I ($n_c \approx 0.02055$), MVtX ($n_c \approx 0.02115$), MVS ($n_c \approx 0.01946$), and OP ($n_c \approx 0.02389$).}
    \label{fig:4}
\end{figure}

In the presence of multipolar textures, the electronic bands undergo folding to the reduced Brillouin zone (RBZ). In Fig.~\ref{fig:4}(a), the folded band structure is depicted for $t'/t=0.8$ and $t''/t=1.225$ in the RBZ with $J_{K_{1,2}}=0$. Figure~\ref{fig:4}(b) shows the band structure in the presence of MSkX-I for the identical hopping parameters, but with $J_{K_1}=J_{K_2}=0.2t$ and $2k_F=\lvert \bm{b}_1 \rvert/6$. The presence of a MSkX-I opens up gaps across the entire RBZ; for instance, notice the opening of gaps at the $M$ and $K$ points for the lowest bands in Fig.~\ref{fig:4}(a) and (b). Meanwhile, the lowest band develops non-trivial Berry curvature [see Fig.~\ref{fig:4}~(c)] and has Chern number $C_n=-1$. The Hall conductance is not quantized, because the system remains gapless in the presence multipolar textures.

Interestingly, in addition to the MSkX phase, the lowest electronic bands also exhibit non-zero Berry curvatures in the RBZ when Kondo-coupled to the MVtX phase, the MVS (MVS') phase, and the OP phase, as illustrated in Fig.~\ref{fig:4}(e)--(f). Due to the spin-orbit coupled nature of the MKLM under consideration, the Berry curvature of the electronic band~\cite{berry1984quantal} gains contributions from both the momentum-space Berry curvature linked to the anomalous current~\cite{jungwirth2002anomalous} induced by the Kondo coupling to the uniform (octupolar) component and the real-space Berry curvature associated with the magnetic flux generated by non-coplanar multipolar textures~\cite{ye1999berry,zhang2020real,verma2022unified}, e.g., the MVtX and MSkX. The latter mirrors the Aharonov-Bohm effect~\cite{aharonov1959} of electrons under a physical external magnetic field in the strong coupling limit.

Figure~\ref{fig:4}(g) shows the Hall conductance $\sigma_{XY}$ within the 2D triangular lattice plane as one tunes through different phases~\cite{supp}.  The MVS (MVS') phase, the MVtX phase, and the MSkX-I phase yield finite $\sigma_{XY}$, consistent with the non-vanishing Berry curvatures of electronic bands in the RBZ. $\sigma_{XY}$ depends on the band structure and the Fermi level's location. Fig.~\ref{fig:4}(h) shows $\sigma_{XY}$ against $J_{K_1}=J_{K_2}=J_K$ while maintaining a constant electron filling fraction within each phase. For the OP phase, $\sigma_{XY}$ remains negligible for $J_K/t<0.3$, gaining significance only for a larger $J_K/t$. It is noteworthy that multipolar phases solely characterized by quadrupolar components, such as the Q-spiral phase and the Q-stripe phase, do not break time-reversal symmetry, resulting in a  zero Hall conductance.

In summary, we present a microscopic mechanism to stabilize topological multipolar textures, such as MSkX, within $f^2$ non-Kramers doublet systems. The critical factor for stabilizing MSkX at zero magnetic field lies in the compass-like interaction on the triangular lattice, induced by bond-dependent RKKY interactions. This bond-dependent interaction is a direct consequence of the spin-orbit intertwined nature inherent in the constituent non-Kramers doublet. Notably, these topological multipolar textures exhibit a large magnitude of Hall conductance of the order of $e^2/h$ even in the \emph{weak coupling limit} [refer to Fig.~\ref{fig:4}(g)(h)]. Consequently, the Hall response of conduction electrons is a valuable tool for detecting these multipolar textures. A family of $\mathrm{Pr}^{3+}$ compounds~\cite{tsujimoto2014heavy,onimaru2012simul,tokunaga2013magnetic,matsumoto2015quadrupole,freyer2018twostage}, particularly in the 2D limit, is a fertile playground to hunt for the MSkX.

We thank A.~Paramekanti, C.~D.~Batista, S.~Banerjee, and Z.~Wang for useful discussions. The work was carried out under the auspices of the U.S. DOE NNSA under contract No. 89233218CNA000001 through the LDRD Program, and was supported by the Center for Nonlinear Studies at LANL, and was performed, in part, at the Center for Integrated Nanotechnologies, an Office of Science User Facility operated for the U.S. DOE Office of Science, under user proposals $\#2018BU0010$ and $\#2018BU0083$. In addition, this research was supported in part by grant NSF PHY-1748958 to the Kavli Institute for Theoretical Physics (KITP).

\bibliographystyle{apsrev4-2}

%

\end{document}


\renewcommand{\figurename}{Fig.~S}
\title{Supplemental Material for ``Multipolar Skyrmion Crystals in Non-Kramers Doublet Systems''}
\author{Hao~Zhang}
\affiliation{Theoretical Division and CNLS, Los Alamos National Laboratory, Los Alamos, New Mexico 87545, USA}
\author{Shi-Zeng~Lin}
\affiliation{Theoretical Division and CNLS, Los Alamos National Laboratory, Los Alamos, New Mexico 87545, USA}
\affiliation{%
Center for Integrated Nanotechnologies (CINT), Los Alamos National Laboratory, Los Alamos, New Mexico 87545, USA
}
\date{\today}
\maketitle

\section{Symmetry analysis for multipolar Kondo couplings}
The derivation of the multipolar Kondo coupling, restricted by local $T_d$ symmetries, is provided in~\cite{patri2020critical}.  Rather than reiterating the procedures for $T_d$, we provide
the recipes to derive the symmetry-allowed Kondo couplings that can be generalized to any arbitrary point group.

As outlined in Ref.~\cite{patri2020critical}, multipolar moments couple to the conduction electron billinears of orbitals and spins. Taking the $e_g$ electrons as an example, the most general Kondo coupling takes the form
\begin{equation}
    \hat{\mathcal{H}}_{K}=\sum_{j}\mathcal{K}^{Ma\alpha}\bigg[\big(\hat{S}_{j}^{M}\big)\otimes\bm{c}_{j}^{\dagger}(\tau^{a}\otimes\sigma^{\alpha})\bm{c}_{j}\bigg],
    \label{eq:general_kondo}
\end{equation}
where $\mathcal{K}^{Ma\alpha}$ denotes the interaction strength that live in the tensor product space of the multipolar operators ($S^{M}$), orbital ($\tau^{a}$), and physical spin ($\sigma^{\alpha}$), and $\bm{c}_{j}$ denotes the spinor of conduction electrons $c_{j,a\alpha}$ in the tensor product space of orbital and spin. The summation goes over all the sites $j$ of the lattice. Note that the tensor product space of multipolar moments, orbital, and physical spin has a dimension of $3\times4\times4=48$ for $e_g$ electrons. Here we have included $\tau_0$ ($\sigma_0$) to denote the orbital (physical spin) singlet.

Now we impose symmetries. In the above-mentioned tensor product space, a symmetry operator $U$ of a point group $\mathcal{G}$ is written as
\begin{equation}
    U=S\otimes\Lambda\otimes\Sigma,
\end{equation}
where $S$ acts on the multipolar moments space, $\Lambda$ acts on the orbital space, and $\Sigma$ acts on the spin space. Let us use $\hat{O}^{Ma\alpha}=\hat{S}^{M}\otimes\bm{c}^{\dagger}(\tau^{a}\otimes\sigma^{\alpha})\bm{c}$ to denote a coupling trilinear. The symmetry-allowed trilinear then must satisfy
\begin{equation}
    \mathcal{K}^{Ma\alpha}\hat{O}^{Ma\alpha}=\mathcal{K}^{Ma\alpha}U\hat{O}^{Ma\alpha}.
\end{equation}
The above constraint is equivalent to a matrix equation $U \mathcal{K} = \mathcal{K}$, where $U$ and $\mathcal{K}$ should now be viewed as a flattened $48\times48$ matrix and a 48-component vector, respectively. Rearranging terms, the matrix equation is written as
\begin{equation}
    (U-\mathbf{I})\mathcal{K}=0,
\end{equation}
where $\mathbf{I}$ is the identity matrix. The set of solutions $\{v_{i}\}$ forms the so-called null space for the operator $F=U-\mathbf{I}$. Let us arrange the columns of $\{v_{i}\}$ into a rectangular matrix $V$. We can then define the projector onto the null space as $P=VV^{\dagger}$. The symmetry-allowed interactions then correspond to the eigenvectors of $P$ with eigenvalue one. For multiple symmetries $\{U_{1},U_{2},\ldots\}\in\mathcal{G}$, we need to calculate the intersection of projectors to null space
\begin{equation}
    P=P_{1}P_{2}\ldots,
\end{equation}
and then find eigenvectors of eigenvalue one. We note that the symmetry analysis discussed above can be generalized to an arbitrary number of orbitals. For instance, considering $p$-orbitals (or $T_{2g}$ orbitals), one just needs to replace the $\tau$ matrices in Eq.~\eqref{eq:general_kondo} by the Gell-Mann matrices.

\section{Variational method}
We applied the variational method outlined in Ref.~\cite{wang_skyrmions_2020} to construct the $T=0$ phase diagrams in the main text. This approach is based on the observation that the magnitude of the ground state ordering wavevector $Q=\lvert \bm{Q}_{\nu} \rvert$ is fixed by $2k_F$ for small conduction electron filling fractions. Subsequently, we set $Q=2k_F=\lvert \bm{b}_1 \rvert/L$, where $L$ is an integer corresponding to the linear size of the super-cell of multipolar spins (pseudo spin-1/2). This super-cell is defined by the basis ${ \bm{a}_1, \bm{a}_2 }$, with $\bm{a}_1$ and $\bm{a}_2$ constituting the basis of the triangular lattice. As noted in~\cite{wang_skyrmions_2020}, the choice of commensurate $2k_F$ is only for convenience of calculation, and any nearby incommensurate $2k_F$ is expected to produce similar phase diagrams.

For a given pair of anisotropic hopping amplitudes ($t'$, $t''$) for conduction electrons, alongside fixed Kondo coupling strengths $J_{K_{1}}=J_{K_{2}}$ and fixed ordering wave vector $Q=2k_F=\lvert \bm{b}_1 \rvert/L$, we calculate the magnetic susceptibility tensor $\chi^{MM'}(\bm{q})$ for all wave vectors within the first reduced Brillouin zone (RBZ) of the $L \times L$ triangular super-cell. Subsequently, we perform a Fourier transform on $\chi^{MM'}(\bm{q})$ to obtain the pseudo spin-1/2 Hamiltonian in real space. Finally, using the open-source semiclassical spin dynamics package Sunny.jl~\cite{Sunny}, we proceed to numerically minimize the classical pseudo spin-1/2 Hamiltonian within the $3L^2$-dimensional phase space of $L \times L$ pseudo spins.

\section{Characterization of other multipolar phases}
\begin{figure}
    \centering
    \includegraphics[width=0.85\columnwidth]{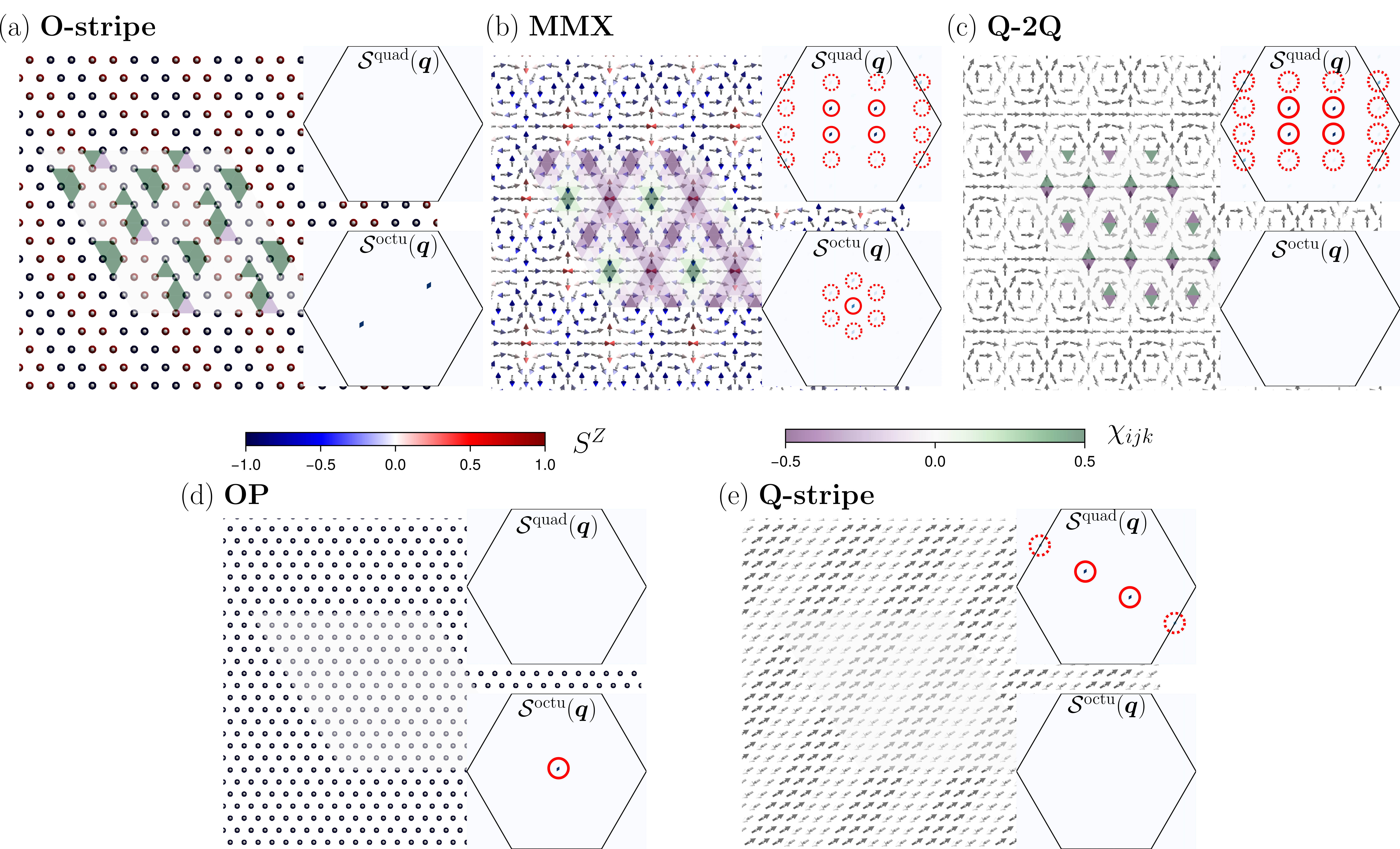}
    \caption{Multipolar (pseudo) spin configurations of multipolar phases. The in-plane components, $S^X$ and $S^Y$, represent the expectation values of two quadrupolar operators, while the out-of-plane component, $S^Z$, corresponds to the expectation of the octupolar component. Insets show the quadrupolar ($\mathcal{S}^{\text{quad}}$) and octupolar ($\mathcal{S}^{\text{octu}}$) structure factors in the first Brillouin Zone (BZ). Solid red circles emphasize dominant peaks, while dotted red circles indicate subdominant peaks.} 
    \label{smfig:1}
\end{figure}

Here, we describe the multipolar phases in Fig.~S~\ref{smfig:1} that have not been addressed in the main text.
\begin{itemize}
\item Octupolar stripe (O-Stripe) phase: In this phase, the octupolar components are organized into a ``up-up-down-down'' stripe-like structure. This phase is discovered exclusively within the short-wavelength phase diagram, where $Q=2k_F=\lvert \bm{b}_1 \rvert/4$.

\item Multipolar meron crystal (MMX) phase: In this phase, the multipolar components are arranged in a meron crystal structure. Within each magnetic unit cell, two merons with a skyrmion charge of 1/2 are present. This phase is observed within a narrow region of the phase diagram, where $Q=2k_F=\lvert \bm{b}_1 \rvert/6$.

\item Quadrupolar double-Q phase (Q-2Q) phase: In this phase, the quadrupolar components adopt an ``in-plane'' double-Q structure.

\item Octupolar polarized (OP) phase: In this phase, the dominant feature is the ferromagnetic alignment of the octupolar components.

\item Quadrupolar stripe (Q-Stripe) phase: In this phase, the quadrupolar components are arranged in a stripe-like structure. This phase is observed for significant large antiferromagnetic compass interactions.
\end{itemize}

\section{Ginzburg-Landau theory}
In clarifying the stabilization mechanism of the multipolar phases reported in the main text, it is noteworthy that, like the dipolar Kondo lattice model~\cite{wang_skyrmions_2020}, the primary function of the long-range RKKY interaction is to induce frustration and introduce a length scale. To substantiate this argument, we employ a process of ``regularizing'' the RKKY interaction, where we consider a short-range $J_{1}$-$J_{2}$ pseudo spin-1/2 model on triangular lattice:
\begin{equation}
    \mathcal{H} = J_{1}\sum_{\langle i,j \rangle} (S_{i}^{x}S_{j}^{x}+S_{i}^{y}S_{j}^{y}
    +\Delta S_{i}^{z}S_{j}^{z})
    +J_{2}\sum_{\langle\langle i,j\rangle\rangle}(S_{i}^{x}S_{j}^{x}+S_{i}^{y}S_{j}^{y}
    +\Delta S_{i}^{z}S_{j}^{z})
    +J_{1}^{\gamma}\sum_{\langle i,j\rangle_{\gamma}}\tilde{S}_{i}^{\gamma}\tilde{S}_{j}^{\gamma}.
    \label{eq:J1-J2}
\end{equation}
The last term accounts for the compass interaction on the nearest-neighbor bonds of the triangular lattice, where $\tilde{S}_i^{\gamma}= \cos\phi_{\gamma} S_i^x + \sin\phi_{\gamma} S_i^y$ defines the pseudo-spin component along the ``compass'' (bond) direction.
Furthermore, we impose $J_{1}<0$ and $J_{2}>0$ to induce frustration. Additionally, as we will soon demonstrate, $J^{\gamma}_1 > 0$ (anti-compass interaction) is necessary to stabilize the multipolar skyrmion crystal (MSkX) phase.

We perform a standard gradient expansion~\cite{lin2016ginzburg,wang2021meron}
\begin{equation}
    \bm{S}_{\bm{r}+\bm{\delta}}=\bm{S}_{\bm{r}} + (\bm{\delta}\cdot\bm{\nabla})\bm{S}_{\bm{r}}
    +\frac{1}{2}(\bm{\delta}\cdot\bm{\nabla})^{2}\bm{S}_{\bm{r}} + \ldots
\end{equation}
to derive the continuum version of Eq.~\eqref{eq:J1-J2} :
\begin{equation}
    \mathcal{H}_{\text{GL}} =\int \mathrm{d}^{2} \bm{r}
    \bigg[-\frac{I_{1}}{2}(\nabla\bm{S}_{\bm{r}})^{2}
    +\frac{I_{2}}{2}(\nabla^{2}\bm{S}_{\bm{r}})^{2}
    +D(S_{\bm{r}}^{z})^{2} 
    -\frac{J_{1}^{\gamma}}{2}\big(\frac{9}{8}\big[(\partial_{x}S_{\bm{r}}^{x})^{2}+(\partial_{y}S_{\bm{r}}^{y})^{2}\big]
    +\frac{3}{4}\big[(\partial_{x}S_{\bm{r}}^{x})(\partial_{y}S_{\bm{r}}^{y})+(\partial_{x}S_{\bm{r}}^{y})(\partial_{y}S_{\bm{r}}^{x})\big]\big)\bigg],
\end{equation}
where
\begin{equation}
    I_{1} = \frac{3}{2}(J_{1}+3J_{2})  \quad 
    I_{2} = \frac{3}{32}(J_{1}+9J_{2}) \quad
    D = 3(\Delta-1)(J_{1}+J_{2}) -\frac{3}{2}J_{1}^{\gamma}.
\end{equation}
Note that we have dropped an overall factor of 1/4 for the pseudo spin-1/2 model under consideration. Henceforth, $\lvert\bm{S}_{\bm{r}}\rvert=1$. By setting $\sqrt{I_{2}/I_{1}}$ as the unit of length and $I_{1}$ as the unit of energy, we obtain the \emph{universal} Ginzburg-Landau (GL) energy functional:
\begin{equation}
    \tilde{\mathcal{H}}_{\text{GL}} = \int\mathrm{d}^{2}\bm{r}
    \bigg[-\frac{1}{2}(\nabla\bm{S}_{\bm{r}})^{2}+\frac{1}{2}(\nabla^{2}\bm{S}_{\bm{r}})^{2}
    +\tilde{D}(S_{\bm{r}}^{z})^{2}
    -\frac{\tilde{J}^{\gamma}}{2}\big(\frac{9}{8}\big[(\partial_{x}S_{\bm{r}}^{x})^{2}+(\partial_{y}S_{\bm{r}}^{y})^{2}\big]+\frac{3}{4}\big[(\partial_{x}S_{\bm{r}}^{x})(\partial_{y}S_{\bm{r}}^{y})+(\partial_{x}S_{\bm{r}}^{y})(\partial_{y}S_{\bm{r}}^{x})\big]\big)\bigg].
    \label{eq:universal}
\end{equation}

The above GL energy functional is universal because all the coupling constants
\begin{equation}
    \tilde{D}\equiv\frac{I_{2}D}{I_{1}^{2}},\quad
    \tilde{J}^{\gamma}\equiv\frac{J_{1}^{\gamma}}{I_{1}},\quad \tilde{\mathcal{H}}_{\text{GL}}\equiv\frac{\mathcal{H}_{\text{GL}}}{I_{1}}.
\end{equation}
are dimensionless. Now we proceed to construct the $T=0$ variational $\tilde{J}^{\gamma}-\tilde{D}$ phase diagram for $\tilde{\mathcal{H}}_{\text{GL}}$.  We follow Refs.~\cite{leonov2015multiply,lin2016ginzburg,wang_skyrmions_2020,wang2021meron} to propose variational ansatzes for several principle multipolar phases that dominate significant regions within the phase diagrams reported in the main text. These ansatzes are derived via Fourier analysis performed on the solutions obtained from \emph{unbiased} calculations, as detailed in the main text.

In contrast to the multipolar phases stabilized by the RKKY interaction, where the magnitude of the ordering wave vector $\lvert \bm{Q} \rvert=2k_F$ is determined by the Fermi wavevector of conduction electrons, $q=\lvert \bm{Q} \rvert$ may vary for different values of $\tilde{J}^{\gamma}$ and $\tilde{D}$ in $\tilde{\mathcal{H}}_{\text{GL}}$. Thus, we introduce $q$ as a variational parameter. Below, $\bm{Q}_1=q\bm{b}_1$, $\bm{Q}_3=q\bm{b}_2$, and $\bm{Q}_2=\bm{Q}_3-\bm{Q}_1$, where $\bm{b}_1$ and $\bm{b}_2$ represent the reciprocal lattice vectors of the triangular lattice.

The octupolar polarized (OP) phase is given by $\bm{S}_{\bm{r}} = (0,0,1)$ with energy density $\tilde{D}$. 

The normalized pseudo spin configuration $\bm{S}_{\bm{r}} = \bm{m}_{\bm{r}} / \lvert \bm{m}_{\bm{r}} \rvert$ for the multipolar vertical spiral (MVS) phase can be parameterized as
\begin{equation}
    m_{\bm{r}}^x = \sin(\bm{Q}_3 \cdot \bm{r}), \quad m_{\bm{r}}^y = 0, \quad m_{\bm{r}}^z = \cos(\bm{Q}_3 \cdot \bm{r}),
\end{equation}
whose GL energy density is $-(4\pi)^2  q^2/6 \left (1 - 4\pi^2 q^2/3 \right) + \tilde{D} \int\mathrm{d}^{2}\bm{r} \cos^2(\bm{Q}_3 \cdot \bm{r})$. For small $\tilde{J}^{\gamma}$, the MVS phase has an instability towards the OP phase as $\tilde{D} \leq -1/4$.

The normalized pseudo spin configuration $\bm{S}_{\bm{r}} = \bm{m}_{\bm{r}} / \lvert \bm{m}_{\bm{r}} \rvert$ for the quadrupolar spiral (Q-spiral) phase can be parameterized as
\begin{equation}
     m_{\bm{r}}^x = \cos(\bm{Q}_3 \cdot \bm{r}), \quad m_{\bm{r}}^y = \sin(\bm{Q}_3 \cdot \bm{r}), \quad m_{\bm{r}}^z = 0.
\end{equation}

The normalized pseudo spin configuration $\bm{S}_{\bm{r}} = \bm{m}_{\bm{r}} / \lvert \bm{m}_{\bm{r}} \rvert$ for the quadrupolar stripe (Q-stripe) phase can be parameterized as
\begin{equation}
     m_{\bm{r}}^x = 0, \quad m_{\bm{r}}^y = \mathrm{sign}[\cos(\bm{Q}_3 \cdot \bm{r})], \quad m_{\bm{r}}^z = 0.
\end{equation}

\begin{figure}
    \centering
    \includegraphics[width=0.5\columnwidth]{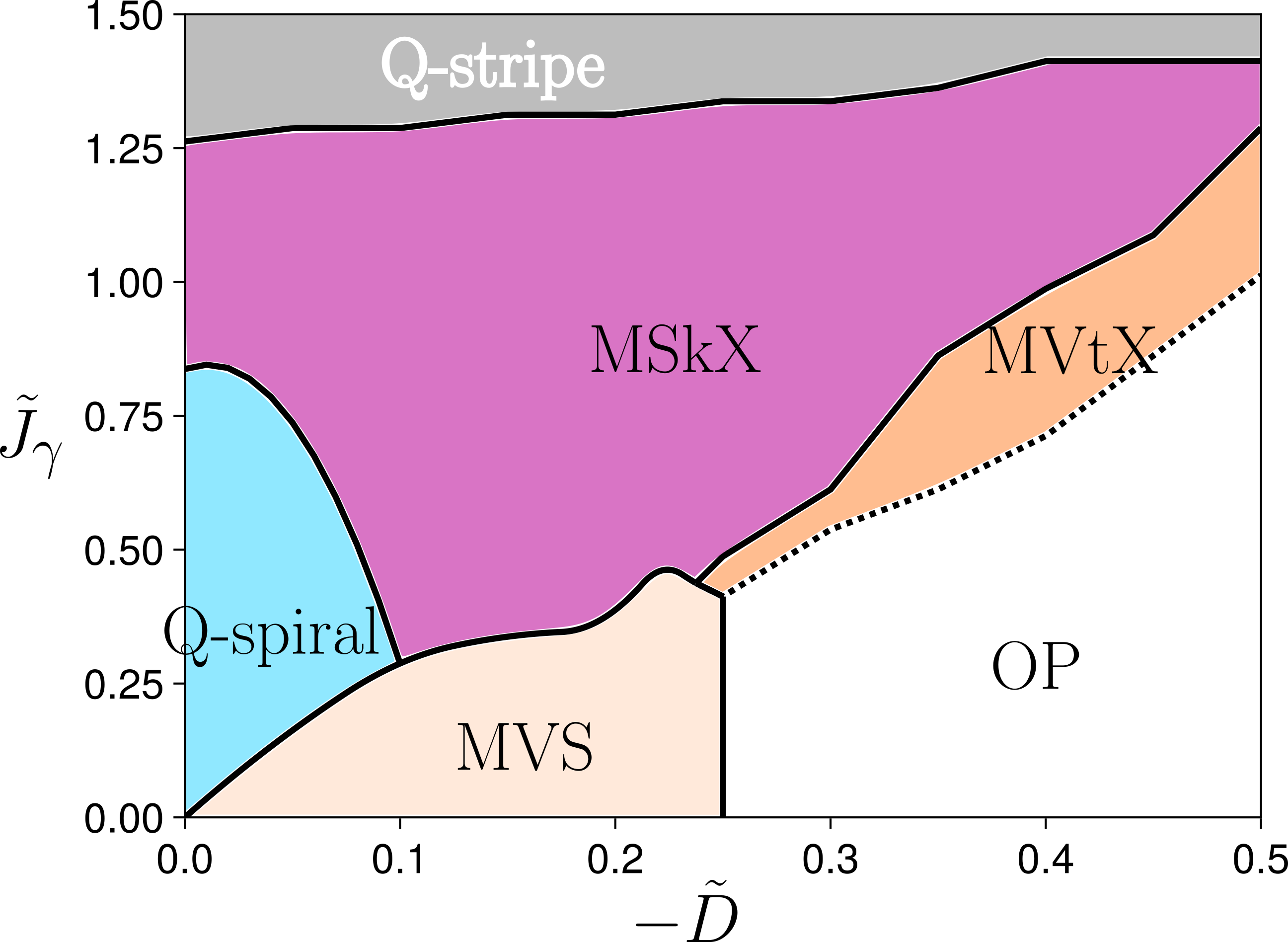}
    \caption{$T=0$ variational phase diagram of the universal Ginzburg-Landau energy function Eq.~\eqref{eq:universal}. Solid lines represent first-order phase boundaries, whereas dotted lines denote second-order phase boundaries.} 
    \label{smfig:2}
\end{figure}

The normalized pseudo spin configuration $\bm{S}_{\bm{r}} = \bm{m}_{\bm{r}} / \lvert \bm{m}_{\bm{r}} \rvert$ for the triple-$\bm{Q}$ phases, including the multipolar vortex crystal (MVtX) state, MSkX-I (skyrmion charge one) state, and MSkX-II (skyrmion charge two) state, can be commonly parameterized as
\begin{align}
    m_{\bm{r}}^x &= a_1 \frac{\sqrt{3}}{2} \left( \sin\left(\bm{Q}_1 \cdot \bm{r} + \frac{3\pi}{4}\right) + \sin\left(\frac{\pi}{2} - \bm{Q}_2 \cdot \bm{r}\right) \right), \nonumber \\
    m_{\bm{r}}^y &= a_1  \left( \sin\left(\bm{Q}_3 \cdot \bm{r} + \frac{3\pi}{4}\right) 
    + \frac{1}{2} \sin\left(\bm{Q}_1 \cdot \bm{r} - \frac{\pi}{4}\right) 
    + \frac{1}{2} \sin\left(\frac{\pi}{2} - \bm{Q}_2 \cdot \bm{r}\right) \right), \nonumber \\
    m_{\bm{r}}^z &= a_0 + a_2 \left( -\cos\left( \bm{Q}_1\cdot \bm{r} + \frac{\pi}{4}\right) 
    + \cos\left( \bm{Q}_2 \cdot \bm{r} \right) 
    - \cos\left( \bm{Q}_3 \cdot \bm{r} + \frac{\pi}{4} \right)\right).
\end{align}

We then optimize different variational ansatzes and find the lowest energy solution, which yields the phase diagram (see Fig.~S~\ref{smfig:2}) for the universal GL energy functional Eq.~\eqref{eq:universal}. Notice that $-\tilde{D}$ ($\tilde{D}>0$) favors finite ``out-of-plane'' (octupolar) pseudo-spin component, resulting in the OP phase for $-\tilde{D} \geq 1/4$. Conversely, positive $\tilde{J}^{\gamma}$ favors a finite ``in-plane'' (quadrupolar) component, leading to the formation of the Q-stripe phase for large $\tilde{J}^{\gamma}$. When $-\tilde{D}$ and $\tilde{J}^{\gamma}$ take small values, the interplay between them gives rise to either an MVS phase or the Q-spiral phase, depending on whether the overall interaction is easy-plane or easy-axis. For intermediate values of $-\tilde{D}$ and $\tilde{J}^{\gamma}$, the ground state is energetically favorable to form triple-$\bm{Q}$ states with winding of spins~\footnote{The energy difference of MSkX-I and MSkX-II is tiny in the GL theory. Hence, we do not distinguish them in the universal phase diagram.}. To understand this, we consider an ansatz for the triple-$\bm{Q}$ state in polar coordinates
\begin{equation}
    \bm{S}(r,\varphi) = \left[\cos\Theta(r) \cos\Phi(\varphi), \cos\Theta(r) \sin\Phi(\varphi),  \sin\Theta(r)\right],
    \label{eq:3Qansatz}
\end{equation}
with $\Phi(r, \varphi) = m \varphi + \varphi_0$, where $m$ is the vorticity (winding number) and $\varphi_0$ is the helicity, which does not impact the GL energy. As pointed out in Ref.~\cite{lin2016ginzburg}, the first two terms of Eq.~\eqref{eq:universal} penalize the triple-$\bm{Q}$ state described by Eq.~\eqref{eq:3Qansatz}. However, a \emph{positive} $\tilde{J}^{\gamma}$ favors the same state with finite vorticity $m$, as the GL energy for the compass term at a particular $(r,\varphi)$ takes the form
\begin{equation}
    -\frac{\tilde{J^{\gamma}}}{2} \frac{9}{8} \frac{1}{r^2} \cos^2[\Theta(r)] \sin^2[\Phi(\varphi)] m^2 .
\end{equation}

\begin{figure}
    \centering
    \includegraphics[width=0.65\columnwidth]{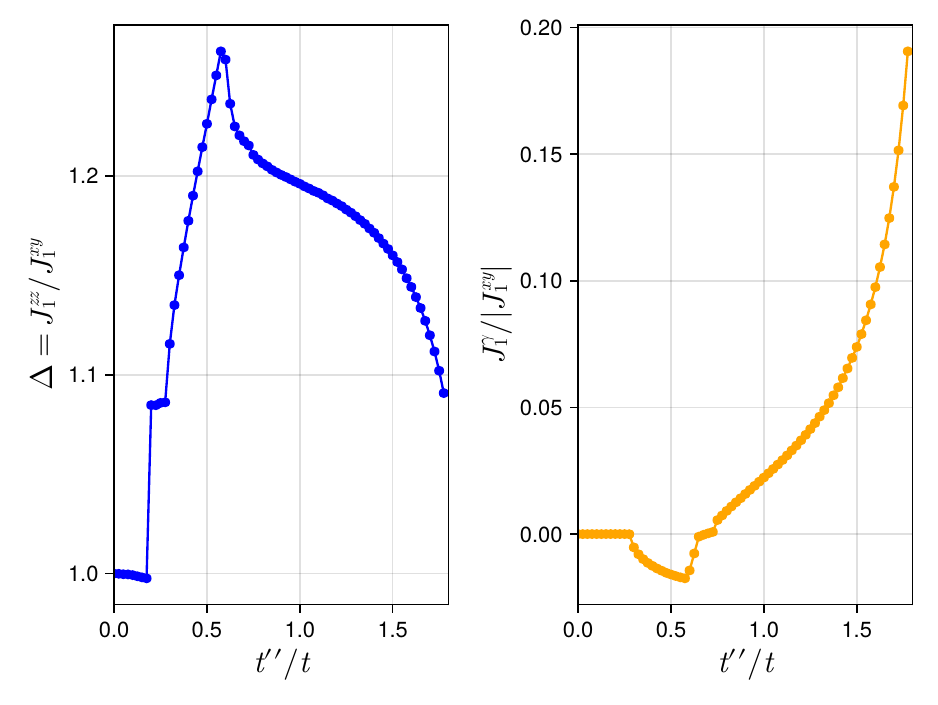}
    \caption{$\Delta = J_{1}^{zz} / J_{1}^{xy}$ and $J_1^{\gamma}/\lvert J_{1}^{xy}\rvert$ as functions of $t''/t$ for $t'=0$ and $2k_F=\lvert \bm{b}_1 \rvert / 8$.} 
    \label{smfig:3}
\end{figure}

Finally, we make a connection between the pseudo-spin-1/2 models discussed in the main text and the GL theory discussed here. To achieve this, we depict $\Delta = J_{1}^{zz} / J_{1}^{xy}$ and $J_1^{\gamma}/\lvert J_{1}^{xy}\rvert$ (for nearest-neighbor bonds with ferromagnetic $J_{1}^{xy} < 0$) as functions of $t''/t$ for $t'=0$ and $2k_F=\lvert \bm{b}_1 \rvert / 8$ in Fig.~S~\ref{smfig:3}. It is noteworthy that similar trends are observed for different $t'$-cuts of $2k_F=\lvert \bm{b}_1 \rvert / 6$ and $2k_F=\lvert \bm{b}_1 \rvert / 8$. The above analysis reveals an alignment between the behaviors depicted in the main text's two mesoscale phase diagrams [refer to Fig.~2(b)(c) of the main text] and the universal phase diagram.

\section{Band structure and Hall conductivity}
The \emph{bare} dispersion relation for the conduction electron hopping Hamiltonian, as described by Eq.~(3) of the main text, is given by
\begin{equation}
    \varepsilon_{\bm{q},\pm} = d_0(\bm{q}) \pm \lvert \bm{d}(\bm{q}) \rvert.
    \label{eq:dispersion}
\end{equation}
Here $\bm{d}(\bm{q})=[d_1(\bm{q}), d_2(\bm{q}), d_3(\bm{q})]$ with components:
\begin{align}
    d_0(\bm{q}) &= -2t   \left( \cos(q_1) + \cos(q_2) + \cos(q_1+q_2) \right), \nonumber \\
    d_1(\bm{q}) &= -2t'' \left( \frac{\sqrt{3}}{2} \left( -\cos(q_2) + \cos(q_1+q_2) \right)\right), \nonumber \\
    d_2(\bm{q}) &=  2t'  \left( \sin(q_1) + \sin(q_2) - \sin(q_1+q_2) \right), \nonumber \\
    d_3(\bm{q}) &= -2t'' \left( \cos(q_1) - \frac{1}{2} \left( \cos(q_2) + \cos(q_1+q_2) \right) \right),
\end{align}
where $(q_1, q_2)$ represent the coordinates of the triangular lattice reciprocal basis vectors $(\bm{b}_1, \bm{b}_2)$. The corresponding eigenvector matrix is given by
\begin{equation}
    \mathcal{U}(\bm{q})=
    \begin{pmatrix}
        \big[d_{3}(\bm{q})+\lvert\bm{d}(\bm{q})\lvert\big]/n_{+}(\bm{q}) & \big[d_{3}(\bm{q})-\lvert\bm{d}(\bm{q})\lvert\big]/n_{-}(\bm{q})\\
        \big[d_{1}(\bm{q})+id_{2}(\bm{q})\big]/n_{+}(\bm{q}) & \big[d_{1}(\bm{q})+id_{2}(\bm{q})\big]/n_{-}(\bm{q})
    \end{pmatrix},
\end{equation}
where
\begin{equation}
    n_{+}=\sqrt{d_{1}^{2}+d_{2}^{2}+(d_{3}(\bm{q})+\lvert\bm{d}(\bm{q})\lvert)^{2}}\quad n_{-}=\sqrt{d_{1}^{2}+d_{2}^{2}+(d_{3}(\bm{q})-\lvert\bm{d}(\bm{q})\lvert)^{2}}.
\end{equation}

When conduction electrons are Kondo coupled to an $L \times L$ array of multipolar spin textures, the size of the Hamiltonian increases to a size of $2L \times 2L$ (accounting for the two $e_g$ orbitals) matrix $\mathcal{H}(\bm{q})$, spanning by the spinor:
\begin{equation}
    c_{\bm{q}, a \alpha} = \frac{1}{\sqrt{N_u}} \sum_{\bm{R}_i} e^{-i \bm{q} \cdot \bm{R}_i} c_{\bm{R}_i, a \alpha},
\end{equation}
where $a$ ($\alpha$) is the sublattice (orbital) index, $N_u$ is the number of magnetic unit cell, and $\bm{R}_i$ is the position of the $i$-th magnetic unit cell. The Hamiltonian is then diagonalized to find the band structure [shown in Fig.~4(b) of the main text] and the eigenvectors:
\begin{equation}
    \mathcal{H}(\bm{q}) \lvert u_{n}(\bm{q}) \rangle = E_n(\bm{q}) \lvert u_{n}(\bm{q}) \rangle.
\end{equation}

Finally, the Hall conductivity is calculated as
\begin{equation}
    \sigma_{xy} = \frac{e^2}{h} \sum_{n} \int_{\mathrm{RBZ}}  \frac{d^2\bm{k}}{(2\pi)^2} f(E_{n}(\bm{k})-E_F) \Omega_{n}(\bm{k}),
\end{equation}
where $f(E_{n}(\bm{k})-E_F)$ is the Fermi distribution function and
\begin{equation}
    \Omega_{n}(\bm{k}) = -2 \Im \sum_{m \neq n} 
    \frac{\langle u_{n}(\bm{k}) \lvert \partial_{k_x} \mathcal{H}(\bm{q}) \lvert  u_{m}(\bm{k}) \rangle
    \langle u_{m}(\bm{k}) \lvert \partial_{k_y} \mathcal{H}(\bm{q}) \lvert u_{n}(\bm{k}) \rangle} {\left(E_n(\bm{k}) - E_m(\bm{k})\right)^2}
\end{equation}
is the Berry curvature of conduction electron in the RBZ.

\bibliographystyle{apsrev4-2}

%